\documentclass[pra,preprint,tightenlines,showpacs]{revtex4}%
\usepackage{amsfonts}
\usepackage{amsmath}
\usepackage{amssymb}
\usepackage{graphicx}%
\begin{document}
\title{Causality and universality in low-energy quantum scattering}
\author{H.-W. Hammer$^{a}$ and Dean Lee$^{b,a}$}
\affiliation{$^{a}$Helmholtz-Institut f\"{u}r Strahlen- und Kernphysik (Theorie) and
\linebreak Bethe Center for Theoretical Physics, Universit\"{a}t Bonn, D-53115
Bonn, Germany\linebreak$^{b}$Department of Physics, North Carolina State
University, Raleigh, NC 27695, USA}
\date{October 8, 2009}

\begin{abstract}
We generalize Wigner's causality bounds and Bethe's integral formula for the
effective range to arbitrary dimension and arbitrary angular momentum.
\ Moreover, we discuss the impact of these constraints on the separation of
low- and high-momentum scales and universality in low-energy quantum scattering.

\end{abstract}
\maketitle
\preprint{HISKP-TH-09-25}

In quantum mechanics causality requires that no scattered wave propagates
before the incident wave first reaches the scatterer. \ This constraint
ensures a scattering amplitude which is analytic in the upper half plane as a
function of energy $E$ \cite{Kronig:1946a, Schuetzer:1951a}. \ For the case of
finite-range interactions the constraints of causality were first investigated
by Wigner \cite{Wigner:1955a}. \ The time delay between the incoming wave and
the scattered outgoing wave can be computed from the energy derivative of the
elastic phase shift, $\Delta t=2\hbar\,d\delta/dE$. \ If $d\delta/dE$ is
negative, this implies a time advance of the outgoing wave. \ However, the
time advance cannot be arbitrarily large since the incoming wave must first
enter the interaction region before the scattered wave can exit. \ Since the
derivative of the phase shift with respect to the energy is involved, this
argument places a bound on the effective range of the scattering amplitude.

Wigner bounds are particularly interesting in the context of low-energy
scattering and universality. \ Universality at low energies arises when there
is a large separation between the short-distance scale of the interaction and
the long-distance scales given by the average particle spacing and thermal
wavelength. \ One example of low-energy universality is the unitarity limit,
which refers to an idealized system where the range of the interaction is zero
and the $S$-wave scattering length is infinite. \ It has been studied most
thoroughly for two-component fermions. \ In nuclear physics, cold dilute
neutron matter is close to the unitarity limit. \ However, most recent
interest in unitarity-limit physics is driven by experiments with cold $^{6}%
$Li and $^{40}$K atoms using magnetically-tuned Feshbach resonances. \ For
reviews of recent cold atom experiments, see Refs.~\cite{Koehler:2006A,
Regal:2006thesis}. \ Theoretical overviews of ultracold Fermi gases and their
numerical simulations are given in \cite{Giorgini:2007a,Lee:2008fa}. \ A
general review of universality at large scattering length can be found in
\cite{Braaten:2004a}.

Several experiments have also investigated strongly-interacting $P$-wave
Feshbach resonances in $^{6}$Li and $^{40}$K \cite{Regal:2003B, Ticknor:2004A,
Zhang:2004A, Schunk:2005A, Gaebler:2007A}. \ An important issue here is
whether the physics of these strongly-interacting $P$-wave systems is
universal, and if so, what are the relevant low-energy parameters. \ A
resolution of these issues would provide a connection between, for example,
the atomic physics of $P$-wave Feshbach resonances and the nuclear physics of
$P$-wave alpha-neutron interactions in halo nuclei. \ Some progress towards
addressing these questions has been made utilizing low-energy models of
$P$-wave atomic interactions
\cite{Ohashi:2004A,Chevy:2004A,Gurarie:2005A,Levinsen:2007A,Gubbels:2007,Jonalasinio:2007A}
and $P$-wave alpha-neutron interactions
\cite{Bertulani:2002sz,Bedaque:2003wa,Nollett:2006su,Higa:2008rx}. A
renormalization group study showed that scattering should be weak in higher
partial waves unless there is a fine tuning of multiple parameters
\cite{Barford:2002je}.

In this letter, we answer the question of universality\ and the constraints of
causality for arbitrary dimension $d$ and arbitrary angular momentum $L$.
\ Our analysis is applicable to any finite-range interaction that is energy
independent, non-singular, and spin independent. \ We present generalizations
of Bethe's integral formula for the effective range \cite{Bethe:1949yr} and
Wigner bounds for arbitrary $d$ and $L$. \ Our results can be viewed as a
generalization of the analysis of Phillips and Cohen \cite{Phillips:1996ae},
who derived a Wigner bound for the $S$-wave effective range for short-range
interactions in three dimensions. \ Here we show that for $2L+d\geq4$, causal
wave propagation produces a fundamental obstruction to reaching the
scale-invariant limit for finite range interactions. \ Instead we find the
emergence of the effective range as a second relevant low-energy parameter
that cannot be tuned to zero without violating causality. \ For the case of
shallow bound states, we show that this second low-energy parameter also
parametrizes the size of the bound-state wavefunction. \ Complementary work
was carried out by Ruiz Arriola and collaborators. A discussion of the Wigner
bound in the context of chiral two-pion exchange can be found in
\cite{PavonValderrama:2005wv} while correlations between the scattering length
and effective range related to the Wigner bound were discussed in
\cite{Cordon:2009pj}.

We consider two non-relativistic spinless particles in $d$ dimensions with a
rotationally-invariant two-body interaction. \ We let $L$ label the absolute
value of the top-level angular momentum quantum number
\cite{Avery:1982,Bolle:1984A}. $\ $For $d\geq2$, $L$ can be any non-negative
integer. \ For $d=1$, the notion of rotational invariance reduces to parity
invariance. \ Here we assume a parity-symmetric interaction and write $L=0$
for even parity and $L=1$ for odd parity. \ We analyze the two-body system in
the center-of-mass frame using units with $\hbar=1$ for convenience. \ With
reduced mass $\mu$ and energy $p^{2}/(2\mu),$ we rescale the radial
wavefunction $R_{L,d}^{(p)}(r)$ as%
\begin{equation}
u_{L,d}^{(p)}(r)=\left(  pr\right)  ^{(d-1)/2}R_{L,d}^{(p)}(r).
\end{equation}
The interaction is assumed to be energy independent and have a finite range
$R$ beyond which the particles are non-interacting. \ Writing the interaction
as a real symmetric operator with kernel $W(r,r^{\prime})$, we have the radial
Schr\"{o}dinger equation%
\begin{align}
p^{2}u_{L,d}^{(p)}(r)  &  =\left[ -\frac{d^{2}}{dr^{2}} +\frac{\left(
2L+d-1\right)  \left(  2L+d-3\right)  }{4r^{2}}\right] \,u_{L,d}%
^{(p)}(r)\nonumber\\
&  +2\mu\int_{0}^{R}dr^{\prime}W(r,r^{\prime})u_{L,d}^{(p)}(r^{\prime
}).\label{radeq}%
\end{align}
The normalization of $u_{L,d}^{(p)}(r)$ is chosen so that for $r\geq R$,%
\begin{align}
u_{L,d}^{(p)}(r)  &  =\sqrt{\frac{pr\pi}{2}}p^{L+d/2-3/2} \left[  \cot
\delta_{L,d}(p)J_{L+d/2-1}(pr)-Y_{L+d/2-1}(pr)\right]  .\label{asymp}%
\end{align}
Here $J_{\alpha}$ and $Y_{\alpha}$ are the Bessel functions of the first and
second kind, and $\delta_{L,d}(p)$ is the phase shift for partial wave $L$.
\ The phase shifts are directly related to the elastic scattering amplitude
$f_{L,d}(p)$, where%
\begin{equation}
f_{L,d}(p)\propto\frac{p^{2L}}{p^{2L+d-2}\cot\delta_{L,d}(p)-ip^{2L+d-2}%
}.\label{amp}%
\end{equation}

In addition to having finite range, we assume also that the interaction is not
too singular at short distances. \ Specifically, we require that the effective
range expansion defined below in Eq.~(\ref{ere}) converges for sufficiently
small $p$ and that $\frac{d}{dr}u_{L,d}^{(p)}$ is finite and $u_{L,d}^{(p)}$
vanishes as $r\rightarrow0$. \ For example, these short-distance regularity
conditions are satisfied for a local potential, $W(r,r^{\prime})=V(r)\delta
(r-r^{\prime})$, provided that $V(r)=\mathcal{O}(r^{-2+\epsilon})$ as
$r\rightarrow0$ for positive $\epsilon$ \cite{Bolle:1984A}. \ In our
discussion, however, we make no assumption that the interactions arise from a
local potential. \ The treatment of spin-dependent interactions with partial
wave mixing is beyond the scope of this analysis. \ For coupled-channel
dynamics without partial wave mixing the analysis can proceed by first
integrating out higher-energy contributions to produce a single-channel
effective interaction. \ In order to satisfy our condition of
energy-independent interactions, this should proceed using a technique such as
the method of unitary transformation described in Ref.~\cite{Fukuda:1954a,
Okubo:1954zz, Epelbaum:1998ka}.

The effective range expansion is%
\begin{align}
&  p^{2L+d-2}\left[  \cot\delta_{L,d}(p)-\delta_{(d\operatorname{mod}%
2),0}\frac{2}{\pi}\ln\left(  p\rho_{L,d}\right)  \right] \nonumber\\
&  =-\frac{1}{a_{L,d}}+\frac{1}{2}r_{L,d}p^{2}+\sum_{n=0}^{\infty}%
(-1)^{n+1}\mathcal{P}_{L,d}^{(n)}p^{2n+4}.\label{ere}%
\end{align}
The term $\delta_{(d\operatorname{mod}2),0}$ is $0$ for odd $d$ and $1$ for
even $d$. \ $a_{L,d}$ is the scattering parameter, $r_{L,d}$ is the effective
range parameter, and $\mathcal{P}_{L,d}^{(n)}$ are the $n^{\text{th}}$-order
shape parameters. $\ \rho_{L,d}$ is an arbitrary length scale that can be
scaled to any nonzero value. \ The rescaling results in a shift of the
dimensionless coefficient of $p^{2L+d-2}$ on the right-hand of Eq.~(\ref{ere}%
), and we define $\bar{\rho}_{L,d}$ as the special value for $\rho_{L,d}$
where this coefficient is zero.

Let $u_{L,d}^{(p)}$ and $u_{L,d}^{(p^{\prime})}$ be radial solutions of the
Schr\"{o}dinger equation for two different momenta. \ We construct the
Wronskian of the two solutions,%
\begin{equation}
u_{L,d}^{(p)}\frac{d}{dr}u_{L,d}^{(p^{\prime})}-u_{L,d}^{(p^{\prime})}\frac
{d}{dr}u_{L,d}^{(p)},\label{Wronksian}%
\end{equation}
and evaluate at some radius $r \geq R$.
Taking the limits
$p^{\prime}\rightarrow0$ and then $p\rightarrow0$, we find that for any $r\geq
R$,%
\begin{equation}
r_{L,d}=b_{L,d}(r)-2\lim_{p\rightarrow0}\int_{0}^{r}dr^{\prime}\left[
u_{L,d}^{(p)}(r^{\prime})\right]  ^{2},\label{r_L}%
\end{equation}
where $b_{L,d}(r)$ is defined as follows. \ For $d=2$ and $d=4$, $b_{L,d}(r)$
can contain logarithmic terms analog to Eq.~(\ref{ere}). \ For the special
case $2L+d=2$, we have%
\begin{equation}
b_{L,d}(r)=\frac{2r^{2}}{\pi}\left\{  \left[  \ln\left(  \frac{r}{2\rho_{L,d}%
}\right)  +\gamma-\frac{1}{2}+\frac{\pi}{2a_{L,d}}\right]  ^{2}+\frac{1}%
{4}\right\}  ,\label{2Ld_2}%
\end{equation}
where $\gamma$ is the Euler-Mascheroni constant and for $2L+d=4$,%
\begin{equation}
b_{L,d}(r)=\frac{4}{\pi}\left[  \ln\left(  \frac{r}{2\rho_{L,d}}\right)
+\gamma\right]  -\frac{4}{a_{L,d}}\left(  \frac{r}{2}\right)  ^{2}+\frac{\pi
}{a_{L,d}^{2}}\left(  \frac{r}{2}\right)  ^{4}\text{.}\label{2Ld_4}%
\end{equation}
For the generic case of $2L+d$ any positive odd integer or any even integer
$\geq6$:%
\begin{align}
b_{L,d}(r) &  =-\frac{2\Gamma(L+\frac{d}{2}-2)\Gamma(L+\frac{d}{2}-1)}{\pi
}\left(  \frac{r}{2}\right)  ^{-2L-d+4}\nonumber\\
&  -\frac{4}{L+\frac{d}{2}-1}\frac{1}{a_{L,d}}\left(  \frac{r}{2}\right)
^{2}\nonumber\\
&  +\frac{2\pi}{\Gamma(L+\frac{d}{2})\Gamma(L+\frac{d}{2}+1)}\frac{1}%
{a_{L,d}^{2}}\left(  \frac{r}{2}\right)  ^{2L+d}.\label{2Ld_6}%
\end{align}
The formula in Eq.~(\ref{2Ld_6}) for $L=0$ in three dimensions was first
derived by Bethe \cite{Bethe:1949yr} and extended by Madsen for general $L$
\cite{Madsen:2002a}.\ \ The results presented here give the generalization to
arbitrary $d$ and arbitrary $L$.

Since the integrand in Eq.~(\ref{r_L}) is positive semi-definite, $r_{L,d}$
satisfies the upper bound%
\begin{equation}
r_{L,d}\leq b_{L,d}(r)\label{upper bound}%
\end{equation}
for any $r\geq R$. \ For $d=3$ our results are equivalent to the causality
bound derived by Wigner \cite{Wigner:1955a}. As noted in the introduction,
the time delay between the incoming wave and the scattered outgoing wave is
proportional to the energy derivative of the elastic phase shift,
$d\delta_{L,d}/dE$. \ Since the incoming wave must first enter the interaction
region before the scattered wave can exit, causality places a bound on
$d\delta_{L,d}/dE$. \ 
The precise quantum mechanical statement of this causality
requirement is that the reciprocal logarithmic derivative $u_{L,d}^{(p)}%
/\frac{d}{dr}u_{L,d}^{(p)}$ has a non-negative energy derivative. \ This fact
can be derived from the Wronskian in Eq.~(\ref{Wronksian}), and a
detailed derivation of this connection will be given in
Ref.~\cite{Hammer:2009a}. \ For finite-range interactions $p^{2L+d-2}%
\cot\delta_{L,d}$ has a convergent effective range expansion, and
$d\delta_{L,d}/dE$ at zero energy is proportional to the effective range
$r_{L,d}$. \ For $d=3$, the Wigner causality bound in zero-energy limit is
equivalent to the bound in Eq.~(\ref{upper bound}) on the effective range.
\ For $S$-wave interactions in three dimensions the upper bound on the
effective range was discussed in Ref.~\cite{Phillips:1996ae}. \ It was
observed that for fixed $a_{L,d}$ the zero-range limit $R\rightarrow0$ is
possible only when $r_{L,d}$ is negative. \ The constraint becomes more severe
for larger $2L+d$. \ For $2L+d\geq4$, the limit $R\rightarrow0$ at fixed
$a_{L,d}$ produces a divergence in the effective range, $r_{L,d}\leq
b_{L,d}(R)\rightarrow-\infty$.

\begin{figure}[ptb]
\centerline{
\includegraphics[
height=3.1in,
]{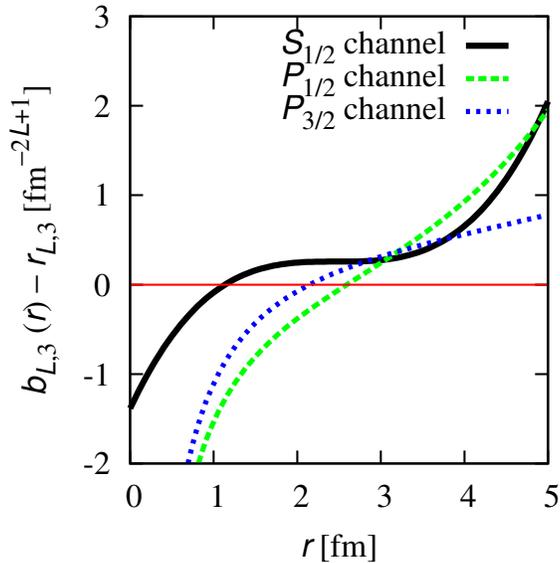}
}\caption{Plot of $b_{L,3}(r)-r_{L,3}$ as a function of $r$ for alpha-neutron
scattering in the $S_{1/2}$, $P_{1/2}$, and $P_{3/2}$ channels. Causality
requires this function to be non-negative for $r \geq R$. }%
\label{alpha_neutron}%
\end{figure}

Our results are exact only for the case where the interaction vanishes for
$r\geq R$. \ For exponentially-bounded interactions of $\mathcal{O}(e^{-r/R})$
at large distances, the results should still be accurate with only
exponentially small corrections. \ For an exponentially-bounded but otherwise
unknown interaction, the non-negativity condition for $b_{L,d}(r)-r_{L,d}$ can
be used to determine the minimum value for $R$ consistent with causality. \ As
an example, we plot $b_{L,3}(r)-r_{L,3}$ for alpha-neutron scattering in
Fig.~\ref{alpha_neutron}. In the plot, we show results for the $S_{1/2}$,
$P_{1/2}$, and $P_{3/2}$ channels. \ We note that a qualitatively similar plot
was introduced for nucleon-nucleon scattering in the $S$-wave spin-singlet
channel \cite{Scaldeferri:1996nx}. The non-negativity condition gives $R\geq1.1$ fm for $S_{1/2}$,
$R\geq2.6$ fm for $P_{1/2}$, and $R\geq2.1$ fm for $P_{3/2}$. \ For
comparison, the alpha root-mean-square radius and pion Compton wavelength are
both about $1.5 $ fm. \ Since the minimum values for $R$ are not small when
compared with these, our analysis suggests some caution when choosing the
cutoff scale for an effective theory of alpha-neutron interactions.

At this point we comment on our requirement that the interactions are energy
independent. \ For energy-dependent interactions it possible to generate any
energy dependence for the elastic phase shifts even when the interaction
$W(r,r^{\prime};E)$ vanishes beyond some finite radius $R$ for all $E$.
\ Under these more general conditions there are no longer any Wigner bounds
and the constraints of causality seem to disappear. \ However, it is
misleading to regard interactions of this more general type as having finite
range. \ As noted in the introduction, the scattering time delay is given by
the energy derivative of the phase shift. \ The energy dependence of the
interaction can by itself generate large negative time delays and thereby
reproduce the scattering of long-range interactions. In this sense the range
of the interaction as observed in scattering is set by the dependence of
$W(r,r^{\prime};E)$ on the radial coordinates $r,r^{\prime}$ as well as the
energy $E$. \ In this case the bound in Eq.~(\ref{upper bound}) can be viewed
as an estimate for the minimum value of this interaction range.

We now consider the scattering amplitude in the low-energy limit
$p\rightarrow0$ while keeping the interaction range $R$ fixed. \ In the
low-energy limit the scattering amplitude depends on just one dimensionful
parameter when $2L+d\leq3$. \ For $2L+d=1$ and $2L+d=3$ the relevant parameter
is $a_{L,d}$, and for $2L+d=2$ it is $\bar{\rho}_{L,d}$. \ When $2L+d\geq4$ a
second dimensionful parameter appears in the non-perturbative low-energy
limit. \ In the limit $\left\vert a_{L,d}\right\vert \rightarrow\infty$, the
upper bounds on the effective range reduce to the form $\bar{\rho}_{L,d}%
\leq\frac{r}{2}e^{\gamma}$ for $2L+d=4$, and%
\begin{equation}
r_{L,d}\leq-\frac{2\Gamma(L+\frac{d}{2}-2)\Gamma(L+\frac{d}{2}-1)}{\pi}\left(
\frac{r}{2}\right)  ^{-2L-d+4}\label{simple_2Ld_6}%
\end{equation}
for $2L+d\geq5$. \ There is no way to suppress the $
p^{2}\ln\left(  p\bar{\rho}_{L,d}\right)  $ term in the effective range
expansion for $2L+d=4$ by fine-tuning parameters because the bound forbids
tuning the argument of the logarithm to 1 as $p\rightarrow0$. \ Similarly, for
$2L+d\geq5$ the upper bound in Eq.~(\ref{simple_2Ld_6}) and the negative
coefficient on the right-hand side prevent setting $r_{L,d}$ to zero to
eliminate the term $\frac{1}{2}r_{L,d}p^{2}$. Hence in both cases we are
left with two relevant parameters in the non-perturbative low-energy limit.
\ This corresponds to two relevant directions near a fixed point of the
renormalization group, and the universal behavior is characterized by two
low-energy parameters. For the case of $P$-wave neutron-alpha scattering in
three dimensions, this issue was already discussed in \cite{Bertulani:2002sz}.
\ Proper renormalization of an effective field theory for $P$-wave scattering
requires the inclusion of field operators for the scattering volume and the
effective range at leading order. \ In the renormalization group study of
\cite{Barford:2002je}, the emergence of multiple relevant directions around 
a fixed point was observed for the repulsive inverse square potential.

For $2L+d\geq4$ this second dimensionful parameter has a simple physical
interpretation for shallow bound states. \ Consider a bound state at
$p=ip_{I}$ in the zero binding-energy limit $p_{I}\rightarrow0^{+}$. \ Let
$P_{>}(r)$ be the probability of finding the constituent particles with
separation larger than $r$:
\begin{equation}
P_{>}(r)=\int_{r}^{\infty}dr^{\prime}\left[  \hat{u}_{L,d}^{(ip_{I}%
)}(r^{\prime})\right]  ^{2},
\end{equation}
where $\hat{u}_{L,d}^{(ip_{I})}$ is the normalized wave function. \ For
$2L+d\leq3$, the probability $P_{>}(r)$ equals $1$ in the limit $p_{I}%
\rightarrow0^{+}$ for any $r$. \ At sufficiently low energies the physics at
short distances is irrelevant, and the bound state wavefunction is spread over
large distances. \ For $2L+d\geq4$, however, the situation is different.
$\ $For $2L+d=4$ the probability is logarithmically dependent on $\bar{\rho
}_{L,d}$ and can be tuned to any value between 0 and 1 \cite{Hammer:2009a}.
\ Similarly for $2L+d\geq5,$%
\begin{equation}
P_{\text{%
$>$%
}}(r)\rightarrow\frac{2\Gamma(L+\frac{d}{2}-2)\Gamma(L+\frac{d}{2}%
-1)}{(-r_{L,d})\pi}\left(  \frac{r}{2}\right)  ^{-2L-d+4}%
\end{equation}
for $r\geq R$. \ For this case the characteristic size of the bound state
wavefunction is $\left(  -r_{L,d}\right)  ^{1/(-2L-d+4)}$.

For $P$-wave Feshbach resonances in alkali atoms our analysis must be modified
to take into account long-range van der Waals interactions of the type
$W(r,r^{\prime})=-C_{6}r^{-6}\delta(r-r^{\prime})$ for $r,r^{\prime}\geq R$.
This raises various new issues such as the applicability of our approach to
power law potentials and the appearance of non-analytic terms in the effective
range expansion. \ These issues will be addressed in detail in
\cite{Hammer:2009a}. \ Here we will briefly discuss the modifications for
potentials with a van der Waals tail in three dimensions only. \ It is
convenient to reexpress $C_{6}$ in terms of the length scale $\beta_{6}=(2\mu
C_{6})^{1/4}$. \ In the following, we set $d=3$ and drop the $d$ subscript.
\ Instead of free Bessel functions, scattering states should be compared with
exact solutions of the attractive $r^{-6}$ potential \cite{Gao:1998a,
Gao:1998b}. \ The effect of the interactions for $r<R$ are described by a
finite-range $K$-matrix $K_{L}(p^{2})$ which is analytic in $p^{2}$
\cite{Gao:2009a}, $K_{L}(p^{2})=\sum_{n=0,1,\cdots}K_{L}^{(2n)}p^{2n}$. \ When
phase shifts are measured relative to free spherical Bessel functions, the
effective range expansion is no longer analytic in $p^{2}$. \ For $L=0$, the
leading non-analytic term is proportional to $p^{3}$. \ For $L=1$ the
non-analytic term is proportional to $p^{1}$, thereby voiding the usual
definition of the effective range parameter.

For a pure van der Waals tail, however, one can still obtain useful
information from our approach. The zero-energy resonance limit is reached by
tuning the lowest-order $K$-matrix coefficient $K_{L}^{(0)}$ to zero. \ It
turns out that for $L=1$ in this limit the $p^{1}$ coefficient in the
effective range expansion also vanishes, and we can therefore define an
effective range parameter for both $S$- and $P$-waves
\cite{Gao:1998b,Flambaum:1999zza},%
\begin{align}
r_{0}  &  =\left[  \Gamma\left(  1/4\right)  \right]  ^{2}\left(  \beta
_{6}+3K_{0}^{(2)}\beta_{6}^{-1}\right)  /(3\pi),\nonumber\\
r_{1}  &  =-36\left[  \Gamma\left(  3/4\right)  \right]  ^{2}\left(  \beta
_{6}^{-1}-5K_{1}^{(2)}\beta_{6}^{-3}\right)  /(5\pi).
\end{align}
For the case of single-channel scattering for alkali atoms, the coefficients
$K_{L}^{(2)}$ are negligible compared with $\beta_{6}^{2}$. \ This is also
true for some multi-channel Feshbach resonance systems \cite{Hanna:2009a}.
\ In these cases we observe that the upper bounds for $r_{L}$ in
Eq.~(\ref{simple_2Ld_6}) are satisfied for $L=0$ and $L=1$ when we naively
take $R\sim\beta_{6}$. \ In general, there may be multi-channel systems where
the coefficients $K_{L}^{(2)}$ cannot be neglected. \ Nevertheless, the
coefficients $K_{L}^{(2)}$ should satisfy Wigner bounds similar to those
derived here for the effective range. \ This may be a useful starting point
for further investigations of multi-channel Feshbach resonances in alkali atoms.

In this letter, we have addressed the question of universality\ and the
constraints of causality for arbitrary dimension $d$ and arbitrary angular
momentum $L$. \ For finite-range interactions we have shown that causal wave
propagation can have significant consequences for low-energy universality and
scale invariance. We find that in certain cases two relevant low-energy
parameters are required in the non-perturbative low-energy limit. \ In the
language of the renormalization group, this corresponds to two relevant
directions in the vicinity of a fixed point. \ In particular, we confirm
earlier findings in the case three dimensions for $P$-wave scattering
\cite{Bertulani:2002sz} based on renormalization arguments and for higher
partial waves in general \cite{Barford:2002je} in the framework of the
renormalization group. \ The analysis presented here concerns only the
question of universality in two-body scattering. \ Universality for higher
few-body systems requires a detailed analysis for each system under
consideration. Effective field theory and renormalization group methods may
again provide a useful starting point here
\cite{Bedaque:1998kg,Barford:2004fz}. Our results may help to
clarify some of the conceptual and calculational issues relevant to few-body
systems for general dimension and angular momentum and their simulation using
short-range interactions.


We are grateful for discussions with D. Phillips and T. Sch\"{a}fer. \ This
research was supported in part by the DFG through SFB/TR 16 \textquotedblleft
Subnuclear structure of matter,\textquotedblright\ the BMBF under contracts
06BN411, 06BN9006, and the US Department of Energy under DE-FG02-03ER41260.

\bibliographystyle{apsrev}
\bibliography{WignerBound_short}

\end{document}